\documentclass[a4paper]{article}

\usepackage{INTERSPEECH2020}
\usepackage{amsmath}
\usepackage{float}
\usepackage{hyperref}
\usepackage{graphicx}
\usepackage{xfrac}
\usepackage{longtable}
\usepackage{threeparttable}
\usepackage{multirow}
\usepackage{enumitem,kantlipsum}

\title{Compact Speaker Embedding: lrx-vector}
\name{Munir Georges$^{1,3}$, Jonathan Huang$^2$\sthanks{*Work done at Intel Labs}, Tobias Bocklet$^{1,4}$}

\address{
  $^1$Intel Labs, Munich, Germany\\
  $^2$Apple Inc, Cupertino, California, USA\\
  $^3$Technische Hochschule Ingolstadt, Germany\\
  $^4$Technische Hochschule N\"urnberg, Germany }
\email{munir.georges@intel.com, jjhuang@apple.com, tobias.bocklet@intel.com}

\begin{document}

\maketitle
 
\begin{abstract}
Deep neural networks (DNN) have recently been widely used in speaker recognition systems, achieving state-of-the-art performance on various benchmarks.
The x-vector architecture is especially popular in this research community, due to its excellent performance and manageable computational complexity.
In this paper, we present the lrx-vector system, which is the low-rank factorized version of the x-vector embedding network.
The primary objective of this topology is to further reduce the memory requirement of the speaker recognition system.
We discuss the deployment of knowledge distillation for training the lrx-vector system and compare against low-rank factorization with SVD.
On the VOiCES 2019 far-field corpus we were able to reduce the weights by 28\% compared to the full-rank x-vector system while keeping the recognition rate constant (1.83\,\% EER).  
\end{abstract}

\noindent\textbf{Index Terms}: speaker recognition, x-vector, low power

\section{Introduction}
\label{INTRODUCTION}
Speaker recognition systems have been popularized in consumer devices such as smart phones and smart speakers, for access control. This is achieved by generating a voice print from the user's speech during interaction with the device and comparing against an existing voice print. Voice prints are usually generated by speaker embeddings of Deep Neural Networks (DNNs), which can also be the underlying feature for diarization in multi-speaker meetings \cite{snyder2019speaker, zhang2019uisrnn}. DNNs have extensively explored in the literature for the generation of speaker embedding with different objective functions \cite{xvector, li2017deep, heigold2016end, Intel:Far:Field:Speaker:Recognition:System:for:VOiCES:Challenge:2019}.  The x-vector system \cite{xvector} emerged as a favorite in the research community, due to its robust training,  state-of-the-art performance and the availability of recipes in the popular Kaldi framework \cite{povey2011kaldi}. Our work here is focused on the x-vector system.

Local inference provides a clear advantages over cloud solution. Examples are: improved protection of user privacy, lower recognition latency, or the autonomy from communications channels. Local inference has been previously addressed for speech recognition  \cite{Accurate:client:server:based:speech:recognition:keeping:personal:data:on:the:client} and spoken language understanding \cite{Speech:Recognition:and:Understanding:on:Hardware:Accelerated:DSP}.  The main challenges in local inference is the limited compute and memory available on the device.  Increasing these requirements have adverse implications to cost and energy efficiency of the device. The memory access operations during inference is identified as a bottleneck for energy efficiency. In this paper, we focus on reducing the memory footprint of an x-vector-based speaker verification system. Furthermore, reduction in memory footprint will lead to lower cost for the device.

There is already a wealth of literature available the focus on the compression of DNNs. Training DNNs with low-rank matrices jointly with the target objective is explored for vision and audio signals, previously.
Novikov et al. \cite{Tensorizing:Neural:Networks} explore low-rank factorization of neural networks using CIFAR-10 and 1000-class ImageNet ILSVRC-2012.
Sak et al. \cite{Long:Short:Term:Memory:Based:Recurrent:Neural:Network:Architectures:for:Large:Vocabulary:Speech:Recognition} use low-rank projection layers in Recurrent Neural Networks (RNN) for speech recognition.
A rank constrained DNN topology for key word spotting is proposed by Nakkiran et al. \cite{Compressing:Deep:Neural:Networks:using:a:Rank:Constrained:Topology}. 
Related work to compression, HashNet uses a low-cost hash function to randomly group weights into hash buckets.
Chen et al. \cite{Compressing:Neural:Networks:with:the:Hashing:Trick} propose and compare this approach with low-rank networks.
Wu et al. \cite{Quantized:Convolutional:Neural:Networks:for:Mobile:Devices} explore quantization of convolutional neural networks and compares them with various alternatives including Low-rank Decomposition and Approximation of Non-linear Responses.
An energy-efficient hardware accelerator using a low-rank approximation is also proposed by Zhu et al. \cite{LRADNN:High:throughput:and:energy:efficient:Deep:Neural:Network:accelerator:using:Low:Rank:Approximation} where inactive neurons are passed by.
Sahraeian et al. \cite{A:study:of:rank:constrained:multilingual:DNNS:for:low:resource:ASR} explore low-rank factorization beyond compression aspects via Singular Value Decomposition (SVD) of the weight matrices to achieve sparse multilingual acoustic models.
More general, Dighe et al. \cite{Low:rank:and:sparse:soft:targets:to:learn:better:DNN:acoustic:models} improve the acoustic model by training using low-rank and sparse soft targets.
Similar success was achieved for Deep Gaussian Conditional Random Fields as explored by, e.g., Chandra et al. \cite{Dense:and:Low:Rank:Gaussian:CRFs:Using:Deep:Embeddings}.
Ding et al. \cite{Deep:Domain:Generalization:With:Structured:Low:Rank:Constraint} describe a structured low-rank constraint using domain-specific and domain-invariant DNNs.
Applying low-rank and low-rank plus diagonal matrix parametrization to RNNs is studied by Barone et al. \cite{Low:rank:passthrough:neural:networks}.
Sharan et al. \cite{Compressed:Factorization:Fast:and:Accurate:Low:Rank:Factorization:of:Compressively:Sensed:Data} explore random projection for low-rank tensor factorization and describe the use on gene expression and EEG time series data. 
Zhang et al. \cite{Structural:sparsification:for:Far:field:Speaker:Recognition:with:GNA} apply structural sparsification on Time-Delay Neural Networks (TDNN) to remove redundant structures.
Alternative approaches are subject to our further research, e.g., binary neural networks as successfully applied to natural language understanding \cite{Ultra:Compact:NLU:Neuronal:Network:Binarization:as:Regularization}.

In this paper, we proposed low-rank x-vector speaker embeddings by deploying knowledge distillation to the training process. We call the resulting embedding lrx-vector.
The paper is organizes as follows:
Section~\ref{sec:xvec} describes the baseline speaker recognition system, with details on the model topology and loss function.  In Section~\ref{sec:Low:Rank:x:vector}, we introduce the modified model topology and the model training methodology using knowledge distillation.  We present the results of our experiments in Section \ref{SEC:EXPERIMENTS} and then conclude the paper in \ref{SEC:CONCLUSION}.
\section{Baseline Speaker Embedding System}
\label{sec:xvec}
\begin{figure}[t]
	\centering
	\vskip+3mm
	\includegraphics[width=15em]{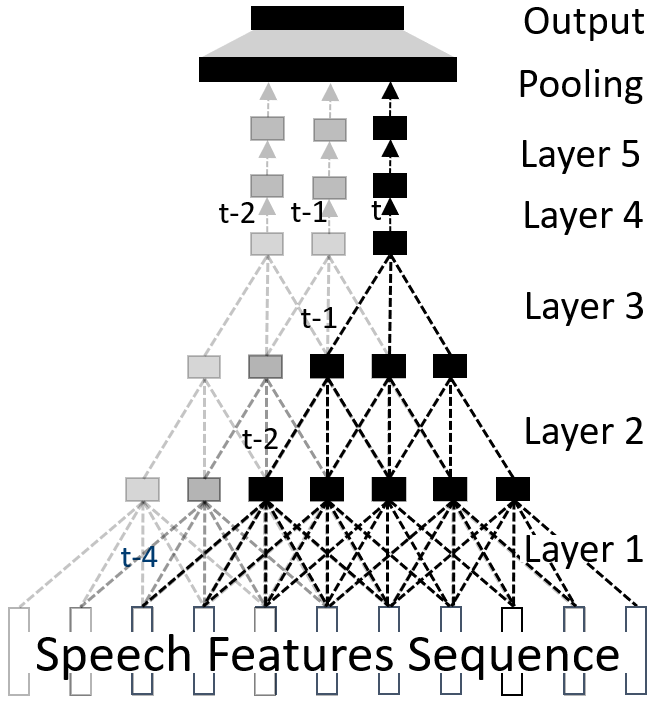}
    \vskip-2mm
	\caption{x-vector speaker embedding. The input sequence of speech features at the top is processed by three TDNN layers. A Statistical Pooling Layer is computed over a speech segment. The Speaker embedding is finally compute by a FFN.}
	\vskip-5mm
	\label{fig:TDNN}
\end{figure}

The x-vector embedding comprises two parts.
First, the feature sequence, i.e., mel-filter bank is processed by layers of TDNNs. 
Second, a statistical pooling layer encodes a segment of speech and computes the embedding vector by a feed forward network.
This architecture is illustrated in Figure \ref{fig:TDNN}.

\subsection{Model topology: x-vector}
\label{ssec:xvec:topology}
This work is based on the x-vector model structure proposed by Snyder et al. \cite{xvector}, with some simplifications.
Compared to the original x-vector model, our architecture, shown on Table \ref{table:xvec_topo}, uses an increased input feature dimension from 24 to 40, reduces the pooling dimension from 1500 to 512 and removes a fully-connected layer between the embedding and speaker output layer. We reduce the embedding dimension from 512 to 256.
In our testing, these modifications did not degrade the recognition performance and result in much lower complexity.
We use this topology as a baseline for comparing against the lrx-vector. 
\begin{table}[H]
\caption{Baseline x-vector configuration for a speech utterance with $T$ frames. Three TDNN, two FFN layer are processing the input sequence. The embedding is computes by a FNN on top of the statistical pooling layer.}
\label{table:xvec_topo}
\small
\centering
\setlength{\tabcolsep}{4.5pt}
\begin{tabular}{|c|c|c|}
    \hline
     & layer context & Affine  \\
    \hline
        Layer1  & [t-2,t+2]     & ($5 \times 40)  \times 512$  \\
        Layer2  & \{t-2,t,t+2\} & ($3 \times 512) \times 512$  \\
        Layer3  & \{t-2,t,t+2\} & ($3 \times 512) \times 512$  \\
        Layer4  & \{t\}         & $512 \times 512$  \\
        Layer5  & \{t\}         & $512 \times 512$  \\
        Stats pooling  & [0,T) & N/A  \\
        Segment & [0,T) & 1024 $\times$ 256  \\
        Output  & [0,T) & 256 $\times$ N \\
    \hline
    \end{tabular}
\end{table}

The output layer is only used during model training; for speaker enrollment and verification, the embedding is taken from the \emph{Segment}-layer (see Table~\ref{table:xvec_topo}).
One speaker embedding is computed for an entire utterance, regardless of its length.
We use the length-normalized cosine distance of this 256-dimension embedding vectors between enrollment and test utterances to produce the speaker recognition score.

\subsection{Loss function}
\label{ssec:loss}
While the conventional softmax loss works reasonably well for training speaker embeddings, it is specifically designed for classification, not verification tasks.
Speaker recognition systems trained with softmax loss typically use PLDA \cite{Ioffe06-PLDA} in the backend to improve separation between speakers.
The triplet loss function, which is designed to reduce intra-speaker and increase inter-speaker distance, has shown to be more effective for speaker recognition \cite{li2017deep}.
Likewise, the end-to-end loss \cite{heigold2016end} has better performance than softmax.
The downside to these kinds of losses is that the training infrastructure is significantly more complex than one used for supervised learning with softmax.
In a prior study ~\cite{Intel:Far:Field:Speaker:Recognition:System:for:VOiCES:Challenge:2019}, we explored the use of several recently proposed loss functions that were first introduced in face recognition research.
These loss functions are drop-in replacements for softmax. Thus, modification to training code is simple with little overhead in training speed.
We found Additive Margin Softmax (AM-softmax) \cite{wang2018additive} to perform best in the far-field test set, and incorporating PLDA did not improve performance against the simpler cosine distance.
The elimination of the PLDA in the inference pipeline makes the entire model easy to deploy to target hardware, with the help of tools such as the Intel\textregistered\space Distribution of OpenVINO\textsuperscript{TM} toolkit\footnote{https://docs.openvinotoolkit.org/}.

\section{Compact Speaker Embedding System}
\label{sec:Low:Rank:x:vector}

\subsection{Model topology: lrx-vector}
\label{ssec:low:rank:TDNN}
First, we analyze the vanilla TDNN layer, which we represent by a feed forward network (FFN).
The layer of the baseline x-vector topology described in Section \ref{ssec:xvec:topology} processes the input $x_t\in \mathbb{R}^{c \times n}$ at time $t$.
It is a concatenation of $c$ feature vectors according to the layer context.
For example, layer 2 of Table \ref{table:xvec_topo} is defined by following row:
\begin{center}
    \begin{tabular}{|c|c|c|}
        \hline
         & layer context & Affine  \\
        \hline
        Layer2  & \{t-2,t,t+2\} & ($3 \times 512) \times 512$  \\
        \hline
    \end{tabular}
\end{center}
Here, $c=3$ for layer context $t-2, t, t+2$ of features generated from layer one at $3$ time steps.
Each input feature to this layer has $n=512$ dimensions.
The overall TDNN output has $m=512$ dimensions. 
The input sequence is shifted by one to process the output for the next time step.
The FFN representing the TDNN layer with weight matrix $\mathbf{W} \in \mathbb{R}^{(c \times n) \times m}$, but no bias is defined as follows:
\begin{align}
    y = \Phi ( \mathbf{W}x_t )
\end{align}
where $\Phi$ a non linear activation function, i.e., ReLU in this paper.
The number of overall weights in a TDNN is ${c \cdot n \cdot m}$.
In the lrx-vector embedding, the TDNN matrix described above is replaced by two matrices ${\mathbf{W_a} \in \mathbb{R}^{(c \times n) \times k}}$ and  ${\mathbf{W_b} \in \mathbb{R}^{k \times m}}$.  The output of the TDNN layer is
\begin{align}
    y = \Phi \left( (\mathbf{W_a} \mathbf{W_b})x_t \right)
\end{align}
where $W_a$ and $W_b$ are low-rank representations of $W$ with  low-rank constant ${1 < k < n}$.
\begin{table}[tb]
    \caption{lrx-vector system configuration for a T-frame speech utterance. Matrices of the four layers with most weights are replaced by low rank matrices.}
    \label{table:lrxvec_topo}
    \small
    \centering
    \setlength{\tabcolsep}{4.5pt}
    \begin{tabular}{|c|c|c|}
        \hline
         & layer context & Affine  \\
        \hline
        Layer1  & [t-2,t+2]     & ($5 \times  40) \times 512$)  \\
        Layer2  & \{t-2,t,t+2\} & ($3 \times 512) \times k_2, k_2 \times 512$  \\
        Layer3  & \{t-2,t,t+2\} & ($3 \times 512) \times k_3, k_3 \times 512$  \\
        Layer4  & \{t\}         & $512 \times k_4, k_4 \times 512$  \\
        Layer5  & \{t\}         & $512 \times k_5, k_5 \times 512$  \\
        Stats pooling  & [0,T) & N/A  \\
        Segment & [0,T) & 1024 $\times$ 256  \\
        Output  & [0,T) & 256 $\times$ N \\
        \hline
    \end{tabular}
\end{table}
The overall number of weights in a low-rank TDNN layer is ${c \cdot k \cdot (n+m)}$ which is significantly smaller compared to vanilla TDNN layer when $k$ is set properly.
The lrx-vector configuration is presented in Table~\ref{table:lrxvec_topo}.

Finally, less weights need to be stored in non volatile memory for lrx-vector system.
On the other hand, the compute increases which is most often not an issue on recent DSP platforms including efficient matrix-matrix multiplication units.

\subsection{Training}
\label{ssec:low:rank:loss:function}
We have explored several different ways of training the lrx-vector system:
\begin{enumerate}[leftmargin=*]
\item \textbf{Random initialization:} The network is initialized by PyTorch's default random initialization.  It is trained in the same way as the baseline x-vector system using the AM-softmax loss described in Section \ref{ssec:loss}. 

\item \textbf{$\text{lrx-SVD}_0$:} A x-vector baseline system is trained, and singular value decomposition (SVD) is performed on the weight matrices.  For each layer, we keep a subset of the singular values.  It is similar to work done by Nakkiran et al \cite{Compressing:Deep:Neural:Networks:using:a:Rank:Constrained:Topology}. The network is not trained any further after SVD.

\item \textbf{$\text{lrx-SVD}_F$:} A network obtained from $\text{SVD}_0$ is fine tuned using the baseline training system until convergence with lowered learning rate.

\item \textbf{Knowledge distillation:} This is a method of using a larger teacher network to train a smaller student network to achieve better performance than it is possible with the smaller network alone. It has been successfully applied in computer vision tasks \cite{Distilling:the:Knowledge:in:a:Neural:Network}. We find the use of a well-trained full-rank x-vector (i.e. the baseline system described in Section \ref{sec:xvec}) as the teacher to the lrx-vector being particularly effective.  Our model training procedure is modified with a  loss function combining contributions from knowledge distillation (KD) and AM-softmax (AMS):
\begin{equation}
\label{eq:kd}
    L_{\text{KD, AMS}} = \alpha L_{\text{KD}} + (1-\alpha) L_{\text{AMS}}
\end{equation}
Here, $L_{KD}$ can be computed by Kullback Leibler divergence (KLD), Mean Square Error (MSE) or Cosine Similarity (COS).  We will make these comparisons in the experiments. Determining the weight $\alpha$ with, e.g., a grid search is time and compute intensive.
We partially circumvent this by applying an idea derived from multi-task learning, previously proposed by Du et al. \cite{Adapting:Auxiliary:Losses:Using:Gradient:Similarity}.
The KD loss gets minimized as long as its gradient has non-negative cosine similarity with the target gradient.
The teacher is ignored, otherwise.
Hence, we minimize following equation with Gradient Cosine Similarity (GCS) to train our speaker identification lrx-vector embedding with $\alpha=0.5$:
\begin{align}
    L_{\text{GCS, KD, AMS}} = \begin{cases}
             L_{\text{KD, AMS}} & \cos{(L_{\text{KD}}, L_{\text{\text{AMS}}})} > 0     \\
             L_{\text{AMS}}   & \text{otherwise}
        \end{cases}   
\end{align}
\end{enumerate}

\section{Experiments}
\label{SEC:EXPERIMENTS}
Our systems were developed by using Voxceleb 1 and 2. The proposed compact speaker identification system is evaluated using the Voxceleb 1 \& 2 data set.
The data is described by McLaren et al. \cite{The:2016:Speakers:in:the:Wild:Speaker:Recognition:Evaluation}, Nagrani et al. \cite{Voxceleb:Large:scale:speaker:verification:in:the:wild}, \cite{VoxCeleb:a:large:scale:speaker:identification:dataset} as training data. Evaluation was performed on the VOiCES challenge far-field text-independent dataset \cite{Voices:Obscured:in:Complex:Environmental:Settings}. The VOiCES development set was used to optimize our system. We present results for the development and evaluation sets using equal error rate (EER) and minimum decision cost function (minDCF) metrics as defined in the VOiCES challenge \cite{nandwana2019voices}.

Our training data is prepared by applying 4x data augmentation. For each augmented speech file we convolve a randomly chosen room impulse response (RIR) from 100 artificially generated by Pyroomacoustics \cite{scheibler2018pyroomacoustics} and 100 selected from Aachen Impulse Response Database \cite{jeub2009binaural}, then mixing in with randomly chosen clips from Google's Audioset under Creative Commons \cite{gemmeke2017audio}. The SNR for mixing was uniformly distributed between 0 and 18\,dB. We extract 40-dimensional mel-filterbank features from 25\,ms frames with 15\,ms overlap.  The features are mean-normalized on a 3-second sliding window.

Our system was developed using the PyTorch\footnote{https://pytorch.org/} framework.  For the baseline system, we used an initial learning rate of 0.1 and decaying to a final learning rate of 0.0001 in 30 epochs of the training data.
Training with knowledge distillation and fine-tuning started at lower learning rate of 0.01.
A weight decay of $1e-6$ is used in all experiments.
We trained all networks until convergence was achieved.
This took most often no longer than 20 epochs. 


\subsection{Low-Rank Factorization}
\begin{figure}[t]
	\centering
	\includegraphics[width=23em]{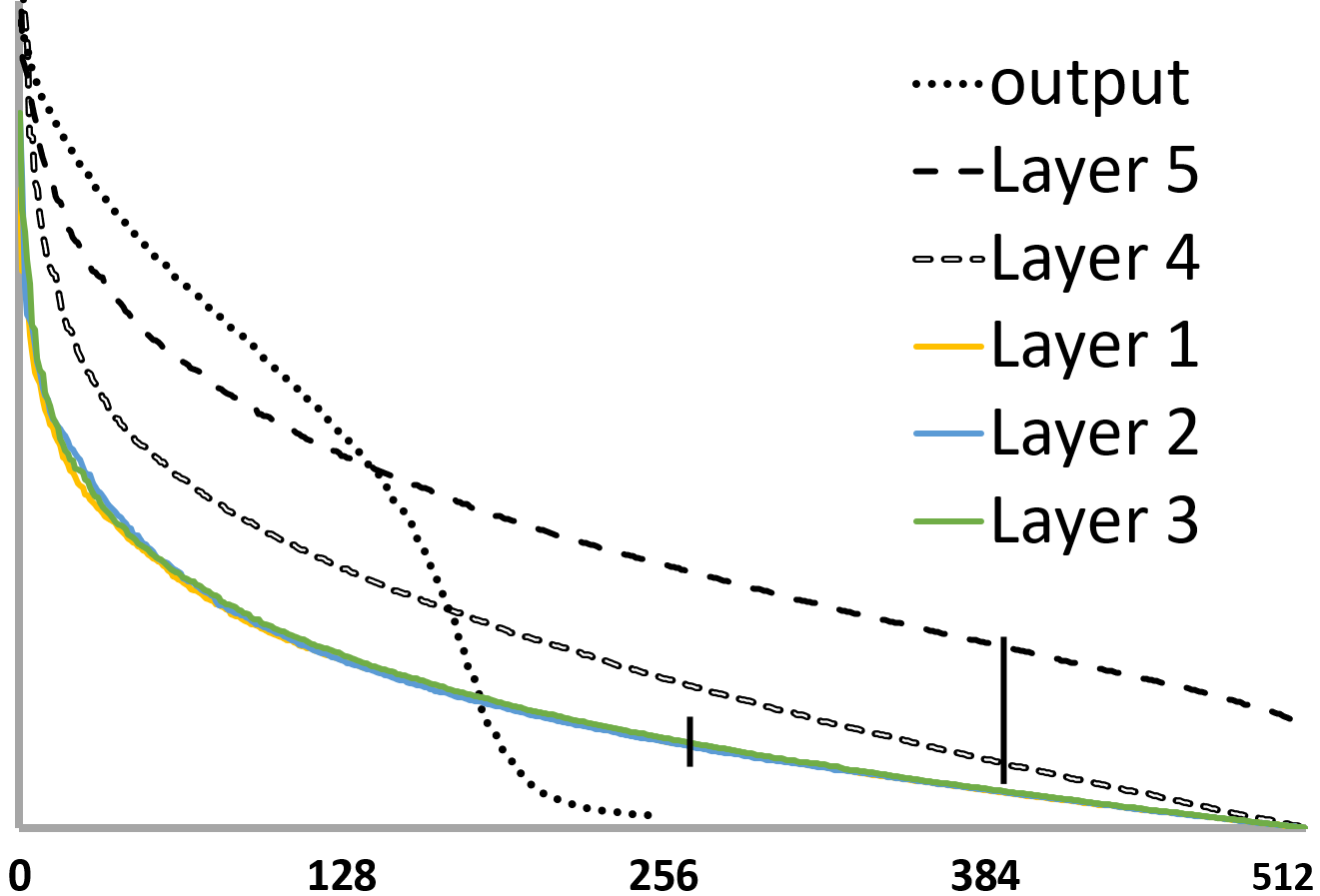}
	\caption{Singular Values of x-vector model matrices. The low-rank constant is, e.g., ${k_2=k_3=256}$ and ${k_4=k_5=384}$.}
	\label{fig:SVD:values}
\end{figure}

An SVD is used to factorize the matrices of a previously trained x-vector system 
Figure \ref{fig:SVD:values} shows the singular values for each layer.
We have empirically figured out that a low rank input layer and low rank layers after the stats pooling significantly decreases the speaker identification accuracy.
In contrast, Nakkiran el al. \cite{Compressing:Deep:Neural:Networks:using:a:Rank:Constrained:Topology} use a low-rank first layer in key-word spotting, successfully.
In this paper, we set ${k_2=0.5 n_2}$, ${k_3=0.5 n_3}$ and ${k_3=0.75 n_3}$, ${k_4=0.75 n_3}$ where $n_2, n_3, n_4$ and $n_5$ are the dimension of the input vectors to layer $2$ to $4$.
It cannot be ruled out that a full grid-search finds better choices for $k_i$ given a target number of overall weights in the lrx-vector.
Automatically determining optimal $k_i$ is subject to our further research.

Table \ref{table:random:vs:svd} compares scaled full-rank x-vector to the randomly initialized lrx-vector, $\text{lrx-SVD}_0$ and $\text{lrx-SVD}_{\text{F}}$.  Note that we linearly scaled the output dimension of each layer in the x-vector for all layer with the same constant factor before the stats pooling, in order to obtain the same model size of $550k$ parameters as in the lrx-vector.  The results indicate that SVD, even with fine-tuning, did not produce better results than random initialization.  Next, we focus the attention to another technique.
\begin{table}[H]
\begin{center}
\caption{Factorization with Singular Value Decomposition (SVD) compared to training from scratch}
\label{table:random:vs:svd}
    \begin{tabular}{ |c|c|c|c|c| } 
         \hline 
         550k& \multicolumn{2}{|c|}{Dev} & \multicolumn{2}{c|}{Eval}\\
         weights & EER & minDCF& EER & minDCF \\
         \hline
            full rank         & 2.78 & 0.307 &	\textbf{6.69} & \textbf{0.483}\\
            lrx-Random init   & \textbf{2.73} &	\textbf{0.289} & 6.76 &	0.484 \\
            $\text{lrx-SVD}_0$    & 3.23  & 0.335  & 7.39 & 0.539 \\
            $\text{lrx-SVD}_{\text{F}}$   & 2.74 & \textbf{0.289} & 6.76 & 0.484 \\
        \hline
    \end{tabular}
\end{center}
	\vskip-5mm
\end{table}


\subsection{Training with Knowledge Distillation}
\label{ssec:knowledge:distillation}
For our research, we choose the teacher to be the baseline system as described in Section\,\ref{sec:xvec}.  The EER and minDCF of the teacher system in Table \ref{table:knowledge:distillation} is a strong baseline, competitive with the top x-vector systems of comparable complexity in the VOiCES Challenge \cite{nandwana2019voices}. We use this baseline system to teach a student  lrx-vector system with $550k$ weights, as described in the previous section.

In the first set of results on Table \ref{table:knowledge:distillation}, we present the experiments using the standard KD learning objective from Eq. \ref{eq:kd}, with $\alpha=0.5$.  It can be seen that KD is generally an improvement over random initialization or  $\text{lrx-SVD}_{\text{F}}$, with KD-MSE performing best with relative EER improvement of 
 $7.19\%$ on the development, and $5.68\%$ on the evaluation sets.  Although this set of results is promising, it might be possible to improve further by tuning $\alpha$.  Practically, doing this by grid search will require long training time. 

\begin{table}[H]
\begin{center}
\caption{Comparison of loss functions between teacher and student at the example of the lrx-vector system with overall 550k weights. The teacher is a baseline system with 4800k weights.}
\label{table:knowledge:distillation}
    \begin{tabular}{ |c|c|c|c|c| } 
         \hline 
         lrx-vector& \multicolumn{2}{|c|}{Dev}& \multicolumn{2}{c|}{Eval}  \\
         550k weights & EER  & minDCF& EER & minDCF  \\
         \hline
            $\text{lrx-SVD}_{\text{F}}$   & 2.74 & 0.289 & 6.76 & 0.484 \\
            \hline
            KD-KLD   & 2.62 &  0.290 & 6.72 & 0.484\\
            KD-MSE   & \textbf{2.58} &  \textbf{0.270} & \textbf{6.31} & \textbf{0.452}\\
            KD-COS   & 2.67 &  0.273 & 6.37 & 0.456\\
        \hline
         \hline
         Teacher     & 1.83 & 0.189 & 5.5 & 0.381\\
        \hline
    \end{tabular}
\end{center}
	\vskip-5mm
\end{table}
Table \ref{table:gradient:cosine:similarity} shows the results of gradient cosine similarity as described in Section\,\ref{ssec:low:rank:loss:function}.  Here, we see that the best performance is achieved with cosine similarity distillation loss, with relative EER improvement of $11.15\%$ for development and $3.74\%$ for evaluation sets over $\text{lrx-SVD}_{\text{F}}$.
\begin{table}[H]
\begin{center}
\caption{Knowledge distillation with positive Gradient Cosine Similarity between teacher and student. Otherwise, only Additive Margin Softmax speaker identification loss.}
    \label{table:gradient:cosine:similarity}
    \begin{tabular}{ |c|c|c|c|c| } 
         \hline 
         lrx-vector& \multicolumn{2}{|c|}{Dev} & \multicolumn{2}{c|}{Eval} \\
         550k weights& EER & MinDCF & EER & MinDCF \\
         \hline
            $\text{lrx-SVD}_{\text{F}}$   & 2.74 & 0.289 & 6.76 & 0.484 \\
        \hline
            KD-GCS-KLD  & 2.45 & 0.270 &	6.77 & 0.487 \\
            KD-GCS-MSE  & \textbf{2.42} & 0.272 &	6.49 & 0.470 \\
            KD-GCS-COS  & 2.47 & \textbf{0.266} & \textbf{6.44} & \textbf{0.458} \\
        \hline
         \hline
         Teacher     & 1.83 & 0.189 & 5.5 & 0.381 \\
        \hline
    \end{tabular}
\end{center}
	\vskip-5mm
\end{table}


\subsection{Scaling lrx-vector}
\label{ssec:lrx:vector:at:different:size}
Comparing x- and lrx-vector based speaker identification of different sizes at approximately the same EER is subject of this evaluation section.
The number of weights in the models were adjusted by changing the dimension of each hidden layers by a multiplication factor $<1$. This factor is the same for all layers in the network up to the stats pooling layer. Automatically determining Layer dependent factors is subject of future research.
lrx-vector as well as x-vector systems were trained by knowledge distillation using KD-GCS-COS.
We selected the best  models given the development set as similar in previous sections.
The results are shown in Table\,\ref{table:sid:sizes}.
A $2.1\%$ EER was achieved with an lrx-Vector system that requires to store $800k$ weights in ROM.
This is $73\%$ of the size to a comparable x-vector system that meets the same EER.
\begin{table}[H]
    \begin{center}
        \caption{Weigh reduction of lrx-vector system compared to x-vector system that achieves same EER.}
        \label{table:sid:sizes}
        \begin{tabular}{ |c|c|c| } 
            \hline 
            EER & \multicolumn{2}{c|}{lrx-vector size}\\
            Dev &  \% of x-vector equivalent & \#weights   \\
            \hline
            4.5 & 100 &  134k \\ 
            2.4 &  79 &  550k \\ 
            2.1 &  73 &  800k \\ 
            1.8 &  72 & 1540k \\ 
            \hline
            \hline
            1.83 & \multicolumn{2}{c|}{X-Vector Teacher with 4800k weights} \\
            \hline
        \end{tabular}
    \end{center}
	\vskip-5mm
\end{table}

\begin{figure}[t]
	\centering
	\includegraphics[width=18em]{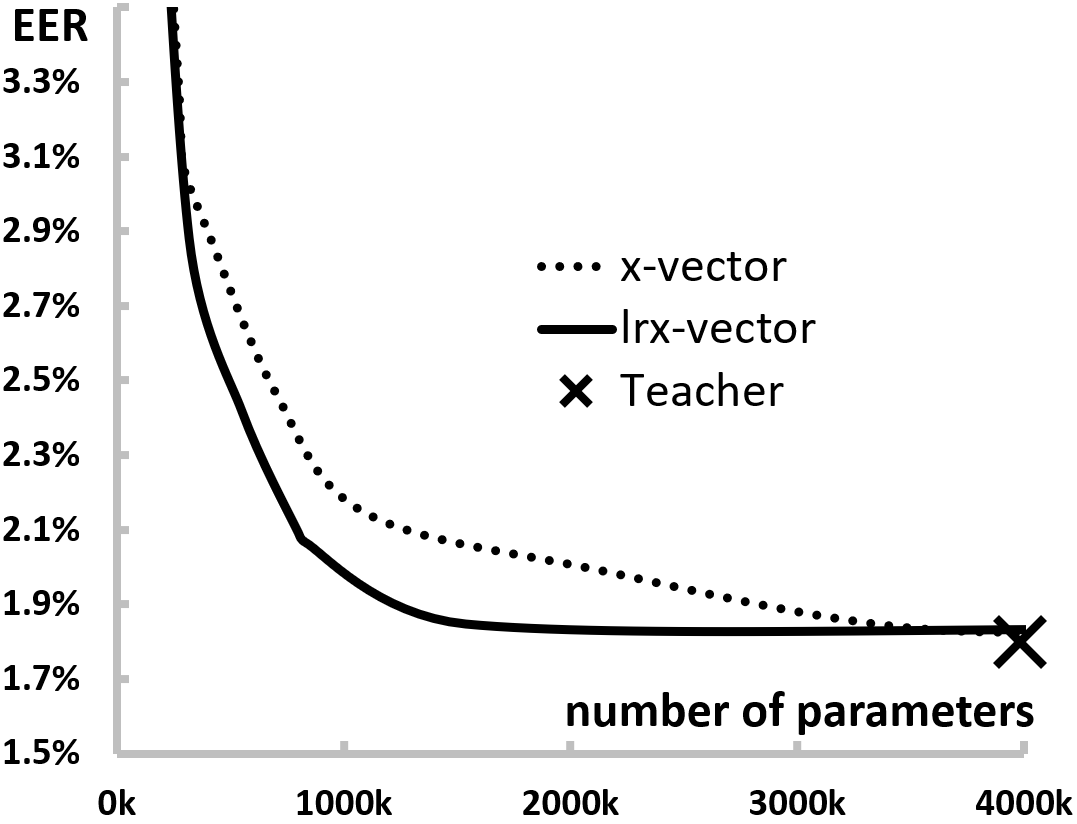}
	\vskip-3mm
	\caption{Achieved Equal Error Rate (EER) on the development set at different number of weights of X- and lrx-Vector Systems.}
	\vskip-5mm
	\label{fig:eer:vs:size}
\end{figure}

Figure \ref{fig:eer:vs:size} illustrates the memory saving of lrx-vector systems compared to x-vector systems at different number of weights.
As expected, there is no improvement over the teacher system when the model is very large, i.e., when the model capacity hits the upper bound of the task.
Very small models do not benefit from low rank matrices, too.
What can be seen is an improvement of the lrx-vector over the x-vector system at common operation points of the system.
In other words, our proposed lrx-vector system shifts the operation point towards smaller models.

\section{Conclusion}
\label{SEC:CONCLUSION}
This paper addresses compact speaker identification by lrx-vector embedding.
We propose a low-rank version of the popular TDNN based x-vector embedding where big matrices are replaced by low-rank matrices.
We address one of the main bottlenecks of low power inference in small edge devices, memory access, by reducing the size of the model.  Using the VOiCES far-field test set, we achieved $28\%$ reduction in the number of parameters compared to the full size model, at the same EER of $1.8\%$.  The lrx-vector is also shown to achieve reduction in model size compared to scaled-down x-vector, at comparable EERs across a wide range of operating points. 
Future research beyond this work can include other means of searching for best knowledge distillation hyper-parameter $\alpha$, and joint low-rank and weight quantization optimizations. 
\bibliographystyle{IEEEtran}
\bibliography{mybib}

\end{document}